\documentstyle[12pt]{article}
\begin{document}
\newcommand{\gx}{\frac{G_{,x}}{G}}
\newcommand{\fx}{\frac{F_{,x}}{F}}
\newcommand{\px}{\frac{P'}{P}}
\newcommand{\nx}{\frac{N'}{N}}
\newcommand{\gt}{\frac{G_{,t}}{G}}
\newcommand{\ft}{\frac{F_{,t}}{F}}
\newcommand{\ftt}{\frac{F_{,tt}}{F}}
\newcommand{\gtt}{\frac{G_{,tt}}{G}}
\newcommand{\fxx}{\frac{F_{,xx}}{F}}
\newcommand{\pxx}{\frac{P''}{P}}
\newcommand{\gxx}{\frac{G_{,xx}}{G}}
\newcommand{\gtx}{\frac{G_{,tx}}{G}}
\newcommand{\dddot}{\mbox{\raisebox{0.5mm}{$\stackrel{...}{M}$}}}

\title{New Non-Separable Diagonal Cosmologies.}

\author{Marc Mars  \thanks{Also at Laboratori de F\'{\i}sica Matem\`atica, 
IEC, Barcelona.} \\
Departament de F\'{\i}sica Fonamental, Universitat de Barcelona, \\
Diagonal 647, 08028 Barcelona, Spain.}
\date{}
\maketitle

\begin{abstract}

We find {\it all} the perfect fluid $G_2$ diagonal cosmologies with the property
that the quotient of the norms of the two orthogonal Killing vectors is
constant along each fluid world-line. We find four different
families depending each one on two or three arbitrary parameters which
satisfy that the metric coefficients are not separable functions. Some
physical properties of these solutions including energy conditions,
kinematical quantities, Petrov
type, the existence and nature of the singularities and whether they contain
Friedman-Robertson-Walker cosmologies
as particular cases are also included.
\end{abstract}


\newpage

\section{Introduction}

The study of exact solutions of Einstein's field equations for a perfect-fluid
energy-momentum tensor is a difficult task even when we assume the existence
of a two-dimensional abelian isometry group acting on the manifold. Indeed,
for these spacetimes the Einstein equations constitute a system of highly
non-linear coupled partial differential equations with two independent
variables which cannot be solved at present in the general case
without making further hypotheses.

As we are interested in studying models which could represent inhomogeneous
cosmologies we will assume that the isometry group two-dimensional
orbits
are spacelike everywhere (the so-called
$G_2$ on $S_2$ metrics) and we will also make the usual hypothesis
that the fluid velocity vector of the perfect fluid is
irrotational. Under these hypotheses, 
the models can be classified {\cite{W1} into four classes depending on
whether the orbits of the group are orthogonally transitive (which means that
the set of two-planes orthogonal at each point to the group orbits are themselves
surface forming) and on the existence of integrable Killing fields.
The simplest (and in fact the most studied up to now) of these classes
is the so-called diagonal cosmologies in which the group orbits are orthogonally
transitive and both the Killing vectors are integrable (which in turn implies
that they can be chosen mutually orthogonal everywhere) so that it can be
seen that there exist coordinates $\{ t,x,y,z \}$ in which the metric
take the Einstein-Rosen's form
\begin{eqnarray}
ds^2 = F(x,t) \left ( -dt^2 + dx^2 \right )
+ G(x,t) P(x,t) dy^2 + \frac{G(x,t)}{P(x,t)} dz^2. \label{met1}
\end{eqnarray}
The first of the known perfect-fluid solutions under these hypotheses
not satisfying an equation of state for a stiff fluid
was found by Wainwright \& Goode {\cite{WG}} some time ago (1980). This family
depends on three arbitrary parameters and the perfect fluid satisfies
a linear equation of state with the pressure proportional to the density.
It has the particularity that the metric coefficient $P$ depends only on
$t$, or equivalently the three-spaces orthogonal to 
the fluid velocity are conformally flat. This solution has also the property
that all the metric functions are separable in the coordinates $t$ and $x$,
that is to say, they are products of one function depending only on $t$ and
another function depending only on $x$. New metrics were
later found in \cite{FS}, \cite{S1} and \cite{Da}, all of them being also
separable metrics with explicit assumptions on the form of the functions
involved. Finally, the case of separable diagonal cosmologies
was extensively treated in a remarkable paper \cite{RS}, where the general
equations for separable cosmologies in co-moving coordinates (not including
the family due to Wainwright \& Goode nor the
solutions satisfying the equation of state for a stiff fluid) were
written and studied. Explicit families were also found showing that
all kinds of singular behaviours were possible even when the
energy conditions were satisfied everywhere (model with a 
big-bang-like singularity and/or big crunch singularity,
metrics with timelike singularities, models with
pressure and density regular everywhere but with singularities in the Weyl
tensor or complete models with no singularities at all). An extensive
discussion of the relation between the existence of non-singular solutions
satisfying energy and causality conditions and the powerful singularity
theorems \cite{HP} was published in \cite{SL}.
More recently,
all the models with pressure equal to density and assuming separation
of variables for diagonal cosmologies have been published in \cite{Go}.

Thus, we see that
the seek of exact solutions for diagonal cosmologies has been performed mainly
assuming the Ansatz of separation of variables in co-moving coordinates. 
However, another approach has also been used in order to find new
diagonal cosmologies. The use of the generalised Kerr-Schild transformation
using Friedman-Robertson-Walker cosmologies as original base metric \cite{SS}
has proven to be fruitful to find new diagonal solutions which are non-separable
in the metric coefficients (some of these metrics have in fact only one Killing
vector field). All of them, however, 
are of Petrov type D everywhere.

Before stating clearly the assumptions we make in this paper, let us note
that any Friedman-Robertson-Walker cosmology (which has a six dimensional
group of isometries acting on three-dimensional spacelike hypersurfaces) 
possesses also a two-dimensional isometry subgroup acting
orthogonally-transitively
on spacelike two-surfaces and with both Killing vectors integrable, so that
the metric can be written in Einstein-Rosen's diagonal form. In fact, these
metrics can be explicitly written as (see for instance \cite{Kr})
\begin{eqnarray*}
ds^2 = R^2(\tau) \left ( -d\tau^2 + d\rho^2 + \Sigma^2(\rho,k) d\phi^2 +
\Sigma'^2(\rho,k) dz^2 \right ),
\end{eqnarray*}
where $R$ is an arbitrary function usually called scale factor,
the function $\Sigma(\rho,k)$ is given by

\vspace{3mm}
\begin{center}
$\Sigma(\rho,k) \equiv \left \{ \begin{array}{c}
                     \sin \rho \hspace{3mm} \mbox{if}\hspace{3mm} k=1  \\
                    \rho \hspace{3mm} \mbox{if}\hspace{3mm} k=0  \\
                    \sinh \rho \hspace{3mm} \mbox{if} \hspace{3mm} k=-1 
                    \end{array} \right. 
$

\vspace{3mm}

\end{center}
and the prime denotes derivative with respect to $\rho$. Thus, for
the Friedman-Robertson-Walker cosmologies the function $P$ in (\ref{met1}) is
given by
\begin{eqnarray*}
P =\frac{\Sigma(\rho,k)}{\Sigma'(\rho,k)},
\end{eqnarray*}
which has the particularity that it does not depend on the coordinate $\tau$.
Thus, the diagonal cosmologies such that the metric coefficient $P$ in
(\ref{met1}) does not depend on $t$ obviously contain
the Friedman-Robertson-Walker as a particular case and therefore it
is worth considering such metrics without assuming other
hypotheses on the form of the metric potentials. In fact, this is the aim of this
paper where we find out all the diagonal $G_2$ cosmologies with
the property that the function $P$ depends only on the spacelike coordinate.
We find five families of exact solutions, each one depending on arbitrary
parameters. However, one of these families is in fact separable so that
we do not study it here. All the other solutions are of Petrov type I and, 
consequently,
they are not included in the non-separable Kerr-Schild families 
in {\cite{SS}} so that they are, to our knowledge, new. For
each family there exist ranges of the parameters such that
the solutions satisfy energy conditions and also three of the four
families contain Friedman-Robertson-Walker models for particular values of the
arbitrary constants. One of the
families represents a stiff fluid and the solution
can be seen to be generated from a vacuum solution using the
procedure due to Wainwright, Ince \& Marshman {\cite{WI}}.

The behaviour of the solutions is analysed and the
existence of singularities and their nature is also studied. One of the 
families is complete and singularity-free, while other two have
a big-bang-like singularity while they are regular in the future and in
the spacelike directions. 

The plan of the paper is as follows. In section 2 we state
the assumptions of this paper and give a coordinate independent
characterisation, we also write the Einstein field equations and find
the general solution of these equations. In section 3, we
write down explicitly all the found solutions and analyse some of their
physical properties, such as the energy density and pressure of the solution,
whether they satisfy energy conditions, the Petrov type, the existence and
nature of the singularities and the kinematical quantities of the fluid
velocity vector.

\section{Field equations and their solutions.}

As stated in the introduction we will seek exact perfect-fluid
solutions of Einstein's field equations with an abelian
two-dimensional group of isometries acting orthogonally
transitively on spacelike two-surfaces and such that
both Killing vectors are integrable. Thus, there exist coordinates
$\{ t,x,y,z \}$ 
in which the metric takes the Einstein-Rosen's form
\begin{eqnarray}
ds^2 = F(x,t) \left ( -\frac{dt^2}{M(t)} + \frac{dx^2}{N(x)} \right )
+ G(x,t) P(x,t) dy^2 + \frac{G(x,t)}{P(x,t)} dz^2. \label{met}
\end{eqnarray}
where the two Killing vector fields are given by
\begin{eqnarray*}
\vec{\xi}= \frac{\partial}{\partial y}, \hspace{1cm}
\vec{\eta}= \frac{\partial}{\partial z}.
\end{eqnarray*}
Note that we could have set $M=1$ and $N=1$ by means of a coordinate
change of the type
\begin{eqnarray*}
t'= t'(t), \hspace{1cm} x'=x'(x),
\end{eqnarray*}
but in some cases is preferable to maintain these two functions in order to
integrate some of the field equations and, in fact, this will be the case
in our situation.

Our objective in this paper is to find out the co-moving perfect-fluid
diagonal cosmologies with the property that the function $P$ of the metric
does not depend on the variable $t$, that is to say, we will assume $P(x)$.
Rewriting this in a coordinate-independent way, we will find 
the perfect-fluid diagonal cosmologies which satisfy
\begin{eqnarray}
\vec{u} \left ( \frac{\vec{\xi_1}\cdot \vec{\xi_1} }{
\vec{\xi_2}\cdot\vec{\xi_2}} \right ) = 0, \label{con}
\end{eqnarray}
where $\vec{u}$ is the fluid velocity vector of the perfect fluid and 
$\vec{\xi_1}$ and $\vec{\xi_2}$ are two arbitrary mutually orthogonal
Killing vector fields. In order to
see the equivalence of the two assumptions, we first note that in a
perfect-fluid solution with a two-dimensional group
of isometries acting orthogonally
transitively on spacelike surfaces, the fluid velocity vector is
necessarily hypersurface orthogonal and, in consequence, there exist coordinates
$\{t,x,y,z 
\}$ in which the metric takes the form (\ref{met}) with the fluid velocity
vector given by
\begin{eqnarray*}
\vec{u} = \sqrt{\frac{M}{F}} \frac{\partial}{\partial t},
\end{eqnarray*}
(co-moving coordinates).
Except in the case that the function $P$ in (\ref{met}) is a constant 
(so that the metric possesses a three-dimensional group of
isometries and is out of the interest of this paper),
it is trivial to see that the only two mutually orthogonal Killing vector fields 
are given by
\begin{eqnarray*}
\vec{\xi_1} = a \frac{\partial}{\partial y}, \hspace{1cm}
\vec{\xi_2} = b \frac{\partial}{\partial z}, 
\end{eqnarray*}
where $a$ and $b$ are arbitrary non-vanishing constants.
The condition (\ref{con}) now reads
\begin{eqnarray*}
\frac{a^2}{b^2} \sqrt{\frac{M}{F}} \frac{\partial}{\partial t} \left ( P^2 \right ) =0
\Longleftrightarrow \partial_t \left (P \right ) = 0,
\end{eqnarray*}
which is exactly the original assumption of $P$ depending only on $x$.

In order to write down the Einstein tensor (and all the other tensor components in
this paper) we will use the orthonormal tetrad given by
\begin{eqnarray}
\mbox{\boldmath$\theta^0$}=\sqrt{\frac{F}{M}}\mbox{\boldmath $dt$},
\hspace{3mm}
\mbox{\boldmath$\theta^1$}=\sqrt{\frac{F}{N}} \mbox{\boldmath $dx$}, 
\hspace{3mm}
\mbox{\boldmath$\theta^2$}=\sqrt{GP} \mbox{\boldmath$dx$}, \hspace{3mm}
\mbox{\boldmath$\theta^3$}=  \sqrt{ \frac{G}{P} }\mbox{\boldmath$dy$}. 
\label{tetr}
\end{eqnarray}
so that the Einstein field equations for a co-moving perfect
fluid energy-momentum tensor read (in units $c=8 \pi G =1$)
\begin{eqnarray*}
S_{00}= \rho, \hspace{1cm} S_{11}=S_{22}=S_{33}= p, \hspace{1cm}
S_{\alpha\beta}=0 \hspace{3mm} (\alpha \neq \beta),
\end{eqnarray*}
where $S_{\alpha\beta}$ ($\alpha=0,1,2,3$) stands fot the Einstein tensor and $\rho$ and $p$ are
the energy density and the pressure of the fluid respectively. Obviously,
there are combinations of these equations involving the Einstein tensor
alone and the calculation of the density and pressure can be performed
once these equations are solved. In our case,
the three equations which are non-identically satisfied are
\begin{eqnarray*}
S_{01}=0, \hspace{1cm} S_{11}-S_{22}=0, \hspace{1cm} S_{22}-S_{33}=0,
\end{eqnarray*}
and they read explicitly
\begin{eqnarray}
S_{01}= \frac{\sqrt{NM}}{2F} \left (\ft \gx +\fx \gt -2 \gtx +  \gx \gt 
\right )= 0, \hspace{28mm} \label{eq01}\\
S_{22}-S_{33} = \frac{N}{F} \left ( - \gx \px - \frac{1}{2} \px \nx - \pxx +
\frac{{P'}^2}{P^2} \right )= 0, \hspace{30mm}  \label{eq2233}\\
S_{11}- \frac{1}{2} \left ( S_{22} + S_{33} \right ) =
\frac{1}{2F} \left ( \ftt M - \frac{F^2_{,t}}{F^2}M + \ft \gt M
+\frac{1}{2} \ft \dot{M} - \fxx N + \frac{F^2_{,x}}{F^2} N  + \right . 
\nonumber\\
\left .
\fx \gx N - \frac{1}{2} \fx N' - \gtt M -\frac{1}{2} \gt \dot{M}
-\gxx N +  \frac{G^2_{,x}}{G^2} N - \frac{1}{2} \gx N' - 
\frac{{P'}^2}{P^2} N \right )= 0,\label{eq1122}
\end{eqnarray}
where the comma means partial derivative
with respect to the variable indicated, the prime stands for ordinary 
derivative with respect to $x$ and the dot indicates ordinary derivative with
respect to $t$.

In order to simplify this system of differential equations, it is 
convenient to define a new function $K(x)$ through
\begin{eqnarray}
\px \equiv \frac{K}{\sqrt{N}}. \label{P}
\end{eqnarray}
We can assume $K$ non-identically vanishing because we are not considering 
solutions with $P$ constant. Thus,
equation (\ref{eq2233}) can be integrated to give
\begin{eqnarray*}
\gx K + K' =0 \Longleftrightarrow G(x,t)= \frac{L(t)}{K(x)},
\end{eqnarray*}
where $L(t)$ is an arbitrary non-vanishing function depending only on the
variable $t$. Substituting this expression for $G$ into the equation
(\ref{eq01}) we get
\begin{eqnarray}
\ft \frac{K'}{K} - \fx \frac{\dot{L}}{L} - \frac{K'}{K} \frac{\dot{L}}{L}= 0.
\label{eq01b}
\end{eqnarray}
In order to handle 
this equation some different cases have to be considered.
First, let us assume $\dot{L}=0$ so that we have
\begin{eqnarray*}
\ft \frac{K'}{K}=0.
\end{eqnarray*}
As $F_{,t}=0$ would imply that no function in the metric (\ref{met}) would depend
on $t$ and consequently $\partial_t$ would be a Killing vector, it follows
necessarily that $K$ is a constant, but in this case the calculation of the density
gives
\begin{eqnarray*}
\rho = - \frac{K^2}{4 F} < 0
\end{eqnarray*}
and this solution is by no means physically reasonable. Thus, we must
restrict our study to the case $\dot{L} \neq 0$. If in addition we had
$K'=0$ equation (\ref{eq01b}) would imply that the function $F$ would not
depend on $x$ and then it can be easily seen that the metric 
(\ref{met}) possesses a third Killing vector field given by
\begin{eqnarray*}
\vec{\xi_3} = 2 \sqrt{N} \frac{\partial}{\partial x} + K \left ( z 
\frac{\partial}{\partial z} - y \frac{\partial}{\partial y} \right )
\end{eqnarray*}
so that we can discard this case. Thus we can assume
$K' \neq 0$ and $\dot{L} \neq 0$
and use the freedom that we still have in the coordinates
$t$ and $x$ to fix these two functions locally to
\begin{eqnarray*}
K = K_0 e^x ,\hspace{1cm} L= K_0 e^t
\end{eqnarray*}
where we have introduced a constant $K_0 \neq 0$ because nothing prevents
the function $K$ to be negative. With this choice, equation
(\ref{eq01b}) takes the simple form
\begin{eqnarray*}
F_{,t} - F_{,x} - F = 0,
\end{eqnarray*}
which can be trivially integrated to give
\begin{eqnarray}
F = e^{t} H(t+x), \label{F}
\end{eqnarray}
where $H$ is an arbitrary non-vanishing function depending only on the
variable $u \equiv t+x$.  Let us define a new function $Z(u)$ by
\begin{eqnarray}
\frac{1}{H} \frac{d H}{d u} \equiv Z \label{H}
\end{eqnarray}
so that (\ref{eq1122}) takes the form
\begin{eqnarray}
\left ( \frac{dZ}{du} + Z \right ) \left (M - N \right ) + \frac{Z}{2} \left (
\dot{M} - N'\right ) + \frac{N'}{2} - K^2  = 0, \label{Zu}
\end{eqnarray}
which is a linear first order ordinary differential equation for the
function $Z$ with
the particularity (which in fact makes it difficult to study) that the 
coefficients do not depend on $u$ but on other coordinates $t$ and $x$.
In order to handle this equation, we can consider $Z$ as a
function of the two variables $t$ and $x$ and impose the
known condition that this function depends only on the variable $t+x$.
Thus, the equivalent system of partial differential equations for the function
$Z(t,x)$ is
\begin{eqnarray}
\left [ \frac{1}{2} \left ( Z_{,t}+ Z_{,x} \right ) + Z
\right ] \left (M-N \right ) + \frac{Z}{2} \left (
\dot{M} - N'\right ) + \frac{N'}{2} - K^2 = 0, \nonumber\\
Z_{,t} - Z_{,x} = 0. \label{sys}
\end{eqnarray}
In order to solve this system, let
us first note that if $M-N$ vanished, we would necessarily have both $M$
and $N$ constant and then the first equation in (\ref{sys}) would be clearly
incompatible. Thus,
we can define a new function $Z_0(t,x)$ by the relation
\begin{eqnarray}
Z \equiv \frac{Z_0 e^{-\left (t+x \right )}}{M-N} \label{Z}
\end{eqnarray}
chosen in such a way that the system (\ref{sys}) takes the simpler form
\begin{eqnarray}
\partial_t Z_0 + \partial_x Z_0 = e^{t+x} \left ( 2 K^2  - N' \right ),
\nonumber
\\
\partial_t Z_0 - \partial_x Z_0 = Z_0 \frac{\dot{M} + N'}{M-N}. 
\label{sys2}
\end{eqnarray}
We can evaluate the integrability
condition $\partial_t \partial_x Z_0 - \partial_x \partial_t Z_0 = 0$ which
gives
\begin{eqnarray}
\left ( 4K^2  - N'' \right ) \left ( M -N \right )
+ \left (2 K^2  - N' \right ) \left (\dot{M} + N' \right )
+ \hspace{5cm} \nonumber \\
\hspace{6cm} 
+ Z_0 e^{- \left ( t+x \right )} \left ( N'' + \ddot{M} + \frac{ N'^2 -
{\dot{M}}^2}{M-N} \right ) = 0, \label{int}
\end{eqnarray}
This equation suggests two possibilities, either with a vanishing or a
nonvanishing coefficient of $Z_0$. In the first case, the following two
equations must be satisfied 
\begin{eqnarray}
\left ( 4K^2  - N'' \right ) \left ( M -N \right )
+ \left (2 K^2  - N' \right ) \left (\dot{M} + N' \right )=0, 
 \nonumber \\
\left ( M- N \right ) \left ( N'' + \ddot{M} \right ) + N'^2 - \dot{M}^2=0. 
\label{sys3} \hspace{13mm}
\end{eqnarray}
In order to find the general solution of this system of two ordinary
differential
equations that mixes functions of $x$ and functions of $t$, we take the
second derivative with respect to $x$ and $t$ of
the second equation to get
\begin{eqnarray*}
\dot{M} N''' - \dddot N' = 0,
\end{eqnarray*}
from what it follows that either $M$ or $N$ are constants or they
satisfy
\begin{eqnarray*}
N''' = b N', \hspace{1cm} \dddot = b \dot{M},
\end{eqnarray*}
where $b$ is an arbitrary constant. Using this information, it is indeed very
easy to find the general solution of (\ref{sys3}). 
There exist two different possibilities given by

\vspace{3mm}

Solution I $ \left \{ \begin{array}{l}
                         N= N_0 + R_0 e^{2 x} \\
                         M = N_0 + S_0 e^{-2 t}
                         \end{array}
                \right .  $ 
                
\vspace{3mm}

Solution II $ \left \{ \begin{array}{l}
                         N= N_0 + K^2_0 e^{2 x} \\
                         M = N_0 + S_0 e^{2 t}
                         \end{array}
                \right . $

\vspace{3mm}

\noindent where $N_0$ ,$R_0$ and $S_0$ are arbitrary constants. With these
explicit forms for the functions $M$ and $N$, the partial differential
equations (\ref{sys2}) can be integrated to give $Z_0$
and therefore we can immediately calculate $Z$ from its definition
(\ref{Z}) and $H$ from (\ref{H}). The results
are given by the following expressions (note that in the Solution I
we have to distinguish between three subcases in order to perform explicitly
the integral for $H$):

\begin{itemize}

\vspace{3mm}

\item Solution I with $ S_0=0$.
 
\vspace{3mm}

\hspace{5mm} $\displaystyle{ Z= \frac{R_0 - K_0^2 - c e^{-2 
\left (t+x\right )}}{2 R_0} \Longrightarrow H= H_0 \exp \left ( \frac{R_0
- K_0^2}{2 R_0} \left (t+x\right) + \frac{c}{4 R_0} e^{-2\left ( t+x \right )} 
\right)} $.

\vspace{3mm}

\item Solution I with $ R_0=0$.
 
\vspace{3mm}

\hspace{17mm} $\displaystyle{ Z= \frac{K_0^2 e^{2 \left ( t+ x
\right )} + c}{2 S_0} \Longrightarrow  H=H_0 \exp \left ( \frac{K_0^2}{4 S_0} 
e^{2\left (t+x \right )} + \frac{c}{2 S_0} \left (t+x \right ) \right )}$.

\vspace{3mm}

\item Solution I with $ R_0\neq 0$ and $S_0\neq0$.
 
\vspace{3mm}

$\displaystyle{Z= \frac{ \left ( K_0^2 - R_0 \right ) e^{2 \left (t + x\right )}
+ \frac{S_0}{R_0} \left ( R_0 - K_0^2 - 4b R_0\right )}{
2S_0 - 2 R_0 e^{2 \left (t+x \right )} } \Longrightarrow }\\
\displaystyle{ \hspace{7cm} 
H= H_0 e^{ \left ( \frac{1}{2} - \frac{K_0^2}{2 R_0} - 2 b
\right ) \left ( t + x \right ) }\left | S_0 - R_0 e^{ 2 \left ( t+ x\right)}
\right |^b}$.
 
\vspace{3mm}

\item Solution II.
 
\vspace{3mm}

\hspace{3cm} $\displaystyle{Z= -2c e^{-2 \left ( t+x \right )}
\Longrightarrow
H = H_0 \exp \left ( c e^{-2 \left ( t +x \right )} \right )}$.

\vspace{3mm}

\end{itemize}
where the arbitrary constants $b$, $c$ and $H_0$ come from the integration of the
differential equations (\ref{Z}) and (\ref{H}).

Thus, we have already found the general solution of the system (\ref{sys2})
in the case that the integrability condition is identically satisfied. We
can now move to the second case when we have
\begin{eqnarray}
N'' + \ddot{M} + \frac{ N'^2 - {\dot{M}}^2}{M-N} \neq 0 . \label{cond}
\end{eqnarray}
Now, the integrability condition (\ref{int}) gives explicitly
$Z_0$ in terms of $M$ and $N$ as
\begin{eqnarray}
Z_0 =  e^{t+x} \left ( M-N \right ) \frac{\left ( N'' - 4 K^2  
\right ) \left ( M -N \right )
+ \left (N' - 2 K^2  \right ) \left (\dot{M} + N' \right )}{
\left ( M -N \right ) \left ( N'' + \ddot{M} \right ) +  N'^2 -\dot{M}^2} ,
\label{Z0}
\end{eqnarray}
and substuting this expression in each of the equations in
(\ref{sys2}) we get two differential relations involving the functions $M$, $N$
and their derivatives. The second of these equations reads explicitly
\begin{eqnarray}
\left ( \dddot\dot{M} - \ddot{M}^2 \right ) A + 
\left ( \dddot M  - \ddot{M} \dot{M} \right ) A' +
\left( \ddot{M} M - \dot{M}^2 \right ) A'' + \dddot \left (
A N' - A' N \right ) + \nonumber\\
\ddot{M} \left (A' N' - A'' N \right ) +
\dot{M} \left ( A' N'' - A N''' \right ) + 4 M K^2  \left (
N''' - 2 N'' \right ) + B = 0, \label{mix}
\end{eqnarray}
where we have set
\begin{eqnarray*}
A \equiv N' - 2 K^2,  \hspace{7mm} B \equiv  2 K^2
\left ( N'''N' - 2 N N''' -{N''}^2 + 2N'N''+ 4 N N'' - 4 {N'}^2 \right ).
\end{eqnarray*}
The first equation in (\ref{sys2}) gives an even more complicated relation
between $M$ and $N$ but we will not need its explicit
expression here.

The study of the differential equation (\ref{mix}) is much more difficult that
in the previous case and in fact some different subcases must be considered
again.
We will not make here the whole detailed study of these possibilities, and we
will only indicate how to handle them in order to write down
explicitly the solutions that
arise and
which also satisfy the first equation in (\ref{sys2}). This is due to the
fact that most of the calculations have been performed with
the aid of computer algebra so that they are impossible to reproduce here.
To begin with, we must
note that if $A$ vanishes identically, we have from (\ref{Z0}) that $Z_0=0$ and 
therefore we also have $Z=0$ and $F= F_0 e^t$, where $F_0$ is an arbitrary
positive constant. It is not difficult to see that in this case, the metric
(\ref{met}) possesses more than two Killing vector fields.
Thus, we can assume that $A$ is non-vanishing and divide equation (\ref{mix})
by $A$ and take the derivative with respect to the variable $x$ so that the term
in $ \left ( \dddot\dot{M} - \ddot{M}^2 \right )$ disappears. Consequently,
the equation
\begin{eqnarray}
\left ( M \dddot - \ddot{M} \dot{M} \right ) \left ( \frac{A'}{A} \right )'
+ \left ( \ddot{M} M - \dot {M}^2 \right ) \left ( \frac{A''}{A} \right )'
+ \dddot \left ( \frac{A N' - A' N}{A} \right )' + \nonumber \\
\ddot{M} \left (
\frac{A' N' - A'' N}{A} \right )' +\dot{M} \left ( \frac{A' N'' - A N'''}{
A} \right )' + M \left (\frac{4 K^2 \left ( N'''-2N''\right )}{A} \right )'
+ \left ( \frac{B}{A} \right )' = 0 \label{mix2}
\end{eqnarray}
must be satisfied. From this equation we learn that either 
\begin{eqnarray*}
\left ( \frac{A'}{A} \right )'=0 \Longleftrightarrow A' = b A, \hspace{1cm}
b= \mbox{ const}
\end{eqnarray*}
or we can divide the equation (\ref{mix2}) by 
$\left (A'/A \right) '$ and perform the derivative with respect to 
$x$ so that the quadratic term $ \left ( \dddot M - \ddot{M} \dot{M}
\right )$ also disappears. It is now obvious that continuing
with this procedure we finally get a linear third order differential equation
for $M$
\begin{eqnarray}
\alpha_1 \dddot + \alpha_2 \ddot{M} + \alpha_3 \dot{M} + \alpha_4 M +
\alpha_5 = 0, \label{mixx}
\end{eqnarray}
where $\alpha_i$, (i=1,\dots,5) are some complicated expressions depending
only on $x$ which contain some derivatives of quotients of the functions
of $x$ appearing in the original equation (\ref{mix}). Except in the particular
case that all the coefficients
$\alpha_i$ in (\ref{mixx}) are identically vanishing it follows that 
$M$ must satisfy a relation of the type
\begin{eqnarray}
\dddot = a_1 \ddot{M} + a_2 \dot{M} + a_3 M + a_4, \label{rel}
\end{eqnarray}
where $a_1, a_2, a_3$ and $a_4$ are now constants. This equation can be 
trivially solved for $M$ although the explicit solution takes different forms
depending on the multiplicity and the real or complex character of the roots
of the polynomial
\begin{eqnarray*}
X^4 - a_1 X^3 - a_2 X^2  -a_3 X - a_4 = 0.
\end{eqnarray*}
Using these explicit forms for the function $M$ some long but straightforward
calculations allow us to restrict the values of $a_1, a_2, a_3$ and $a_4$
which make the system (\ref{sys2}) compatible and then find
the solution for $N$. Considering also the particular
cases that we have been letting aside in the way to the equation (\ref{mixx}) and which
are not contained in (\ref{rel}) we finally get 
the following set of solutions which constitute the general solution
of the system (\ref{sys2}) satisfying the condition (\ref{cond})

\vspace{3mm}

Solution III $ \left \{ \begin{array}{l}
                         N= N_0 \\
                         M = N_0 + S_0 e^{-2 t} + S_1 e^{-6 t}
                         \end{array}
                \right . $ 

\vspace{3mm}

Solution IV $ \left \{ \begin{array}{l}
                         N= N_0 + R_0 e^{-6x} \\
                         M = N_0 + S_0 e^{-2 t}
                         \end{array}
                \right . $
                
\vspace{3mm}
                               
Solution V $ \left \{ \begin{array}{l}
                         N= N_0 + e^{2x} \left ( R_0 e^{cx} - \frac{4 K_0^2}{
			c} 
                         - K_0^2 \right ) \\
                         M = N_0 + S_1 e^{-2 t}
                         \end{array}
                \right . $ 
                
\vspace{3mm}
         
\noindent where the constants $N_0$, $R_0$ and $S_1$ are arbitrary and the
remaining constants $S_0$ and $c$ are restricted to be non-vanishing.

Once the explicit forms the the metric functions $M$ and $N$ are known, we can
substitute in (\ref{Z0}) and back into (\ref{Z}) to find the expressions for
the functions $Z(t+x)$. Performing the integration of (\ref{H}) we
finally find the function $H(t+x)$. The results are trivially found to be

\vspace{3mm}

$\displaystyle{
\begin{array}{l}
\displaystyle{\mbox{Solution III} \hspace{1cm} Z= \frac{K_0^2}{2S_0} e^{2 \left (t+x\right)}
\Longrightarrow H = H_0 \exp \left( \frac{K_0^2}{4S_0} e^{2 \left (t+x\right)} 
\right )}. \\

\vspace{3mm}

\displaystyle{\mbox{Solution IV} \hspace{1cm} Z=\frac{3 S_0 + K_0^2 e^{2 \left (t+x
\right )}}{2 S_0} \Longrightarrow H = H_0 e^{\frac{3}{2} \left (t+x
\right )} \exp \left( \frac{K_0^2}{4S_0} e^{2\left ( t+x \right )} \right)}.\\

\vspace{3mm}

\displaystyle{ \mbox{Solution V} \hspace{1cm} Z= \frac{c+2}{c+4} \Longrightarrow
H = H_0 e^{\frac{c+2}{c+4} \left (t+x\right )}}.
\end{array} }$

\vspace{3mm}

From the expression for $H$ in the solution V we note that
all the metric coefficients $F$, $G$ and $P$ are separable functions of the
coordinates $t$ and $x$ and therefore this solution is contained in the general
case of diagonal cosmologies separable in co-moving coordinates which was
exhaustively studied by Ruiz\&Senovilla \cite{RS}. Thus we will not study this
solution here any further and we concentrate in the
four families I, II, III and IV.

\section{Properties of the Solutions.}

In this section we will list all the solutions that we have found and
give some of their physical properties such as the energy density and pressure
of the perfect fluid, the kinematical quantities of the fluid velocity vector
and the Petrov type. We will also discuss whether
the solutions contain Friedman-Robertson-Walker models as
particular subcases, if
the energy conditions are satisfied for some ranges of the values of
the parameters and the existence of singularities and their nature.

For each of the four families of the solutions found in the previous section
(Solutions I, II, III and IV) we have already written the functions $M(t)$,
$N(x)$ anf $H(t+x)$, while the metric coefficient $G(x,t)$ is common for all
the cases and is given by
\begin{eqnarray*}
G = e^{t-x}. 
\end{eqnarray*}
In order to get all the metric components we still have to determine 
$F(x,t)$ and $P(x)$. The first of these
functions is trivially
obtained from $H(x+t)$ using (\ref{F}) and $P(x)$ is found
by solving the equation (\ref{P}) which reduces to a quadrature.

For the
general metric (\ref{met}) with $P$ depending only on $x$ the non-vanishing
components (in the tetrad (\ref{tetr})) of the kinematical quantities for
the fluid velocity are given by the following expressions

\vspace{3mm}

$\begin{array}{ll}
\mbox{\bf Expansion} & 
\displaystyle{ \theta = \frac{\sqrt{M}}{\sqrt{F}} \left ( \ft + \gt \right )} \\
& \\
\mbox{\bf Shear tensor} &  
\displaystyle{ \sigma_{11} = \frac{1}{3} \frac{\sqrt{M}}{\sqrt{F}}
\left ( \ft - \gt \right ) 
\hspace{1cm} \sigma_{22}= \sigma_{33} = -\frac{1}{2} \sigma_{11} } \\
& \\
\mbox{\bf Acceleration} & 
\displaystyle{  \mbox{\boldmath$a$}= a_x \mbox{\boldmath $dx$} = 
\frac{1}{2} \fx \mbox{\boldmath$dx$} }
\end{array}$

\vspace{3mm}

\noindent while the vorticity obviously vanishes. The explicit form
for the function $G$  and the fact that $F$ depends on
the variable $t+x$ except for a global factor $e^{t}$ imply the relation  
\begin{eqnarray*}
\fx = \ft - \gt.
\end{eqnarray*}
In consequence, it is enough to list the 
acceleration and the
expansion of the fluid for each solution in order to know all the kinematical
quantities because the shear tensor components can be evaluated from
\begin{eqnarray}
\sigma_{22}= \sigma_{33} = - \frac{1}{2}\sigma_{11}= -\frac{1}{3} 
\frac{\sqrt{M}}{\sqrt{F}} a_x. \label{shear}
\end{eqnarray}
Before listing the possible non-separable families of solutions under the
assumptions of this paper, let us note that some of the constants in
$M$ , $N$ and $H$ above can be reabsorved by means of
a linear coordinate change of the type
\begin{eqnarray*}
& &x = aX + x_0, \nonumber \\
& &t = aT + t_0. \label{cha}
\end{eqnarray*}
In the list that follows, this freedom has always been used
to absorve all the spurious constants so that the parameters appearing in the
solution are relevant (if ocasionally some irrelevant constants still remain,
we explicitly state this fact as well as the reasons why this is so).

\vspace{3mm}

{\bf Solution I} 

\vspace{3mm}

As indicated in the previous section we have to distinguish between three subcases
depending on whether some of the constants $R_0$ or $S_0$ vanish or not. 

\vspace{3mm}

$\displaystyle{ \mbox{\boldmath $ R_0=0$}}$

\vspace{3mm}

The complete set of metric potentials for this solution is given by

\begin{center}

\vspace{3mm}

$\displaystyle{ N=1, \hspace{1cm} M = 1+ \epsilon e^{-2at}, \hspace{1cm}
F = \exp \left ( at + \beta a \left (t+x \right ) + \epsilon c^2
e^{2a \left (t+x \right )} \right )},$

\vspace{3mm}

$\displaystyle{ G= e^{a \left (t-x \right )}, \hspace{1cm} P=\exp \left ( 2c 
e^{ax}
\right )},$

\vspace{3mm}

\end{center}
where $\epsilon$ is a sign. In this family, however, not all the constants
$a$, $\beta$ and $c$ are true parameters identifying a particular solution.
In fact, a change of the type (\ref{cha}) allows us to fix 
$a$ to any desired value whenever $\beta$ is non-vanishing,
while $a$ is a true parameter when $\beta=0$. However,
we prefer to maintain the constants as they stand in order to study the cases
$\beta=0$ and $\beta \neq 0$ together.

The energy density and pressure are given by 
\begin{eqnarray*}
p = \rho = \frac{\epsilon a^2 \left ( 2 \beta + 3 \right )}{4 \exp \left (
3at+ \beta a \left ( t+x \right ) + \epsilon c^2 e^{2a \left (t+x \right )}
\right )}
\end{eqnarray*}
so that the perfect fluid satisfies a stiff equation of state.
The energy (and pressure) is non-negative everywhere if and only
if $\beta$ and $\epsilon$ satisfy
\begin{eqnarray*}
\epsilon \left ( 2 \beta + 3 \right ) \geq 0.
\end{eqnarray*}
The kinematical quantitites of the fluid velocity vector are 
\begin{eqnarray*}
\theta = \frac{ a \sqrt{ 1+ \epsilon e^{-2at}} \left ( 2 \epsilon c^2 e^{2a
\left ( t+x \right )} + \beta + 3 \right )}{ 2 \exp \left ( \frac{a}{2} t
+ \frac{\beta}{2} a \left (t+x \right ) + \epsilon \frac{c^2}{2} e^{2a \left (
t+x \right )} \right )}, \hspace{1cm}
\mbox{\boldmath$a$}= \frac{a}{2} \left ( 2 \epsilon c^2 e^{2a \left ( t+x
\right )} + \beta \right ) \mbox{\boldmath$dx$}
\end{eqnarray*}
The Petrov type is I except for the following particular cases
\begin{itemize}
\item when $c=0$ and $\beta \neq 0$ the solution is type D everywhere.
\item when $c=0$, $\beta=0$ the solution degenerates to conformally flat. 
Thus, the metric represents a Friedman-Robertson-Walker cosmology with
density and pressure (we choose $\epsilon = +1$ so that the energy is
positive) given by
\begin{eqnarray*}
p=\rho = \frac{3}{4} a^2 e^{-3at}.
\end{eqnarray*}
\end{itemize}

\vspace{3mm}

$ \displaystyle{\mbox{\boldmath$S_0= 0$}}$

\vspace{3mm}

The list of metric coefficients is 

\vspace{3mm}

\begin{center}

$ \displaystyle{ N = 1 + \epsilon e^{2 a x}, \hspace{1cm} M = 1, \hspace{1cm}
F = e^{at} \exp \left ( \frac{a}{2}\left ( 1 -\epsilon \beta^2 \right )
\left (t +x \right )
+ c e^{-2a \left (t+x \right )} \right ) },$

\vspace{3mm}

$ \displaystyle{ G = e^{a \left (t-x \right)}, \hspace{1cm} P = 
\left \{ \begin{array}{l} 
            \left ( e^{ax} +  \sqrt{1 + e^{2ax}} \right )^{\beta} \hspace{3mm}
            \mbox{if }  \epsilon = +1\\
           \exp \left ( \beta
          \arcsin \left ( e^{ax} \right ) \right) 
          \hspace{3mm} \mbox{if }  \epsilon = -1 
          \end{array}
 \right .
} $

\vspace{3mm}

\end{center}
where $\epsilon$ is a sign and the same considerations made in the previous
case regarding the parameters $a$, $\beta$ and $c$ hold; when $3-\epsilon
\beta^2 = 0$ the constants $a$ and $c$ identify uniquely a particular
solution in the family while when $3- \epsilon \beta^2 \neq 0$, the constant
$a$ can be fixed to any desired value by means of the coordinate change
(\ref{cha}).

The energy density and pressure read

\vspace{3mm}

\begin{center}

$ \displaystyle{ p= \rho = \frac{\epsilon a^2 c}{ \exp \left ( 
3 a t +
\frac{a}{2} \left ( 1- \epsilon \beta^2 \right ) \left ( t+ x \right ) +
 c e^{-2a \left (t
+x \right )} \right ) } } $

\vspace{3mm}

\end{center}
so that again the perfect fluid obeys a stiff equation of state.
The density (and consequently the pressure) is non-negative in the whole
spacetime 
whenever we have
\begin{eqnarray*}
\epsilon c \geq 0.
\end{eqnarray*}
The kinematical quantities of the fluid velocity vector take the form

\vspace{3mm}

\begin{center}

$ \displaystyle{\theta =  \frac{ a \left (
7 - \epsilon \beta^2 - 4 
 e^{-2a \left ( t+x \right )}\right ) }{ 4  \exp \left ( \frac {1}{2}  
 a t +
\frac{a}{4} \left ( 1 - \epsilon \beta^2 \right ) \left ( t+ x \right ) + 
\frac{c}{2}
e^{-2a \left (t +x \right )} \right )} ,
\hspace{3mm} \mbox{\boldmath$a$}=
\frac{a}{4} \left ( 1  - \epsilon \beta^2 - 4  c e^{-2a \left ( t+x 
\right )} \right ) \mbox{\boldmath$dx$}, } $

\end{center}
and the Petrov type is I everywhere except for the following particular cases: 
\begin{itemize}
\item either when $\beta = 0$ or 
when $1- \epsilon \beta^2 = 0$ and $c \neq 0$ or when
$9-\epsilon \beta^2=0$ and $c=0$ the metric is Petrov type D everywhere.
\item  when $1- \epsilon \beta^2 = 0$ and $c=0$ the solution is conformally
flat. In this case the density vanishes
and the solution is in fact the Minkowski spacetime.
\end{itemize}

\vspace{3mm}

$ \displaystyle{\mbox{\boldmath$S_0 \neq 0$ and \boldmath$R_0 \neq 0$} }$

\vspace{3mm}

The metric coeficients for this solution can be written in the following form

\begin{center}

$\displaystyle{N = \sigma + \epsilon_1 e^{2ax}, \hspace{1cm} M = \sigma + 
\epsilon_2 e^{-2at},
\hspace{1cm} F =  e^{at + a \left (\frac{1}{2} - 2 b - \epsilon_1 \frac{c^2}{2} \right )
\left (t+x \right ) }  \left | e^{2a\left (t+x \right )} - \epsilon_1 \epsilon_2
\right |^b },$

\vspace{3mm}

$\displaystyle{G= e^{a\left (t-x \right)}, \hspace{1cm} P =
         \left \{
         \begin{array}{l}
         \left (e^{ax} + \sqrt{\sigma  + e^{2ax}} \right )^{c}\hspace{3mm} 
         \epsilon_1 = +1\\
         \exp \left ( c \arcsin \left ( e^{ax}
         \right ) \right ) \hspace{3mm} \epsilon_1 = -1 
         \end{array}
         \right . 
}$

\vspace{3mm}

\end{center}
where both $\epsilon_1$ and $\epsilon_2$ are signs and $\sigma$ can take
the values $+1$, $-1$ or $0$ (obviously when any of the two signs $\epsilon_1$
or $\epsilon_2$ are $-1$ then we necessarily have $\sigma=+1$).
The density and pressure are  
\begin{center}

\vspace{3mm}

$\displaystyle{p= \rho= \frac{a^2 \epsilon_2 \left ( 4 - 4b  - \epsilon_1 c^2
\right )}{
4 \left | e^{2a\left (t+x\right)} - \epsilon_1 \epsilon_2 \right |^b \exp 
\left (
\frac{7}{2} a t + \frac{1}{2} a x - a \left (2b + \epsilon_1 \frac{c^2}{2} \right )
\left (t+x \right)
\right )}},$

\vspace{3mm}

\end{center}
which satisfy the energy
conditions whenever the constants are restricted to
\begin{eqnarray*}
\epsilon_2 \left ( 4 - 4b  - \epsilon_1 c^2\right )  \geq 0.
\end{eqnarray*}
The kinematical quantities read
\begin{center}

\vspace{3mm}

$\displaystyle{
\theta = \frac{ a \sqrt{\sigma + \epsilon_2 e^{-2at}}  \left ( e^{2a \left (t+x
\right )} \left [ 7 - \epsilon_1 c^2 \right ] + \epsilon_2 c^2 - 7 
\epsilon_2 \epsilon_1 +
4  b \epsilon_1 \epsilon_2 \right )}
{ 4 \exp \left ( \frac{3}{4} at + \frac{1}{4} ax - a \left (b + \epsilon_1 \frac{c^2}{4}
\right  ) \left ( t+x \right )\right ) \left | e^{2a \left ( t+x \right )} - 
\epsilon_1\epsilon_2 \right |^{\frac{b}{2}} \left ( e^{2a \left (t+x \right )} - 
\epsilon_1 \epsilon_2 \right )} },$

\vspace{3mm}

$\displaystyle{ \mbox {\boldmath$a$}= \frac{a \left ( e^{2a \left ( t+x
\right )} \left [ 1 - \epsilon_1 c^2 \right ] + \epsilon_2 c^2 - \epsilon_1
\epsilon_2 + 
4 b \epsilon_1 \epsilon_2
\right )}{4 \left ( e^{2a \left ( t+x \right )} - \epsilon_1 \epsilon_2 \right )}
\mbox {\boldmath$dx$} }$
 
\end{center}

\vspace{3mm}
\noindent and the Petrov type is again I except
for the following particular values of the constants
\begin{itemize}
\item when $c=0$ the solution degenerates to type D everywhere.
\item when $\epsilon_1 c^2 = 1$ and $b=0$ the solution is conformally flat
and, therefore, it represents a Friedman-Robertson-Walker
cosmology with density (and pressure) given by (choosing $\epsilon_2=+1$ so that
the energy conditions are fulfilled)
\begin{eqnarray*}
p= \rho = a^2 e^{-3at}.
\end{eqnarray*}

\end{itemize}

\vspace{3mm}

Before finishing with the study of this Solution I
let us indicate that all these $p=\rho$
solutions can be seen to be generated from vacuum solutions
using the procedure due to Wainwright, Ince \& Marshman \cite{WI}.
However, as far as we know, they were
previously unknown.

\vspace{3mm}

{\bf Solution II}

\vspace{3mm}

The metric coefficients are

\vspace{3mm}

\begin{center}

$\displaystyle{ M = \alpha + \epsilon e^{2at}, \hspace{1cm} N = \alpha+ e^{2
a x},
 \hspace{1cm} 
F =  \exp \left (at + c e^{-2a \left ( t+x \right)} \right ),}$ 

\vspace{3mm}

$\displaystyle{ G = e^{a\left (t-x\right )}, \hspace{1cm} P = e^{a x} +
\sqrt{e^{2ax} + \alpha},
 }
$

\vspace{3mm}

\end{center}
where $\epsilon$ is a sign and the coordinate change (\ref{cha}) still allows
us to fix the constant $\alpha$ equal to $\pm 1$ (depending on its sign) when
$\alpha \neq0 $ or set $a$ equal to any desired
value when $\alpha=0$. We allow this spurious constant in the family in order
to consider the two subcases together.

The expresions for the energy density and pressure read

\begin{center}

\vspace{3mm}

$ \displaystyle{ \rho = \frac{a^2 \left ( 3 \epsilon  e^{4at+2ax} - 4 \epsilon 
 c
e^{2at} + 4c e^{2ax} \right )}{ 4 \exp \left ( 3at + 2ax + c e^{-2a \left (t
+x \right)} \right ) }, \hspace{1cm}
p = \rho - \frac{2a^2\epsilon  e^{at}}{\exp \left (c e^{-2a \left ( t+x
\right) } \right )} }$

\vspace{3mm}

\end{center}
so that the condition for positive energy density is, obviously
\begin{eqnarray}
3 \epsilon  e^{4at+2ax} - 4 \epsilon c e^{2at} + 4c e^{2ax} \geq 0.
\label{rela+}
\end{eqnarray}
In order to determine the ranges for the parameters $\alpha$ and $c$
which make this condition possible everywhere we must 
distinguish between two cases depending on the sign of $\alpha$.
If $\alpha < 0$ it follows from the positivity of $M$ that $\epsilon = 1$
and then the allowed ranges for the
coordinates $t$ and $x$ are (we are choosing the constant
$a$ positive without loss of generality because a global change
of sign in the coordinates $t$ and $x$ also changes the sign of $a$) 
\begin{eqnarray*}
t > \frac{1}{2a} \log \left ( - \alpha \right ), \hspace{1cm}
x > \frac{1}{2a} \log \left ( - \alpha \right ).
\end{eqnarray*}
The condition (\ref{rela+}) is fulfilled everywhere in this
region whenever we have
\begin{eqnarray*}
\epsilon = 1, \hspace{1cm} c \geq - \frac{3 \alpha^2}{4}, \hspace{1cm}
\alpha < 0.
\end{eqnarray*}
However, from the expression for the pressure it follows that 
there always exist regions of the spacetime which do not fulfil the
condition
\begin{eqnarray*}
\rho + p \geq 0
\end{eqnarray*}
so that we will no longer consider this subcase as physically admissible.
 
When $\alpha \geq 0$, the coordinate $x$ can take arbitrarily
large negative values  and
the dominant term in the relation (\ref{rela+}) when $x \rightarrow 
- \infty$ implies $\epsilon c \leq 0$.
If $\epsilon = +1$ the coordinate $t$ can also take arbitrarily large negative
values and then (\ref{rela+}) in the limit $t \rightarrow - \infty$
clearly shows that the
only possibility is $c=0$. On the other hand, if $\epsilon = -1$, the
expression for $M$ implies that $\alpha$ must be strictly positive
so that we can use the freedom mentioned above to fix $\alpha=1$ and then
the ranges for the coordinates $t$ and $x$ are given by
\begin{eqnarray*}
t < 0 ,\hspace{1cm} - \infty < x < + \infty.
\end{eqnarray*}
It is not difficult to see that (\ref{rela+}) is
fulfilled in this region whenever $c$ is restricted to 
\begin{eqnarray*}
\epsilon = -1 \hspace{1cm} c \geq \frac{3}{4}. \hspace{1cm}
\end{eqnarray*}
The expression for the pressure shows that it is always bigger
than the density and therefore positive everywhere. Thus, the only
energy condition which is not fulfilled is $p \leq \rho$. 

It can be easily seen that when we approach $t=0$ along any timelike curve,
the proper time remains finite. Consequently, either the spacetime is
singular at $t=0$ or it is extendible and our coordinate system only
covers a portion of the whole inextendible spacetime. An analisys of the
curvature invariants near $t=0$ shows that all of them remain finite so that
this solution is likely to be extendible. The same consideration holds when 
$x \rightarrow + \infty$ where the proper length remains finite
and all the curvature invariants have also a finite limit there. In order to
show that this metric is in fact extendible, let us perform the change of
variables
\begin{eqnarray*}
t = - \frac{1}{a} \log\left [ \cosh \left (aT\right ) \right ]
\hspace{1cm}
x = - \frac{1}{a} \log \left [  \sinh\left (a r\right ) \right ]
\end{eqnarray*}
where the range of variation of the new coordinates is given by 
\begin{eqnarray*}
T < 0,  \hspace{1cm} r > 0.
\end{eqnarray*}
The metric in this new coordinates take the following simple form
\begin{eqnarray*}
ds^2 = \frac{1}{\cosh (2aT)} \left [
e^{ c \cosh^2 \left (2aT \right ) \sinh^2 \left (2ar \right )} \left ( - dT^2 +
dr^2 \right ) + \cosh^2 \left ( ar \right )  dz^2 + 
\frac{ \sinh^2
\left (ar \right )}{a^2} d\phi^2 \right ],
\end{eqnarray*}
where we have redefined $a \rightarrow 2a$ and we have renamed $z \rightarrow \phi$ and $y \rightarrow z$ because
this metric is cylindrically simmetric with a regular axis of symmetry at
$r=0$ where the so-called elementary flatness condition is satisfied.
Thus, we need
not extend the coordinate $r$ as it represents a radial coordinate and the
metric is already complete in the spacelike coordinates. Regarding the coordinate $T$,
nothing prevents us to extend its variation range to the whole real line so that
we have in fact extended the original spacetime which is included as the
$T<0$ region.
This solution is
inextendible and the density and pressure in the new coordinates read
\begin{eqnarray*}
\rho = \frac{a^2 \left ( 4 c\cosh^4 (2aT) + 4 c \cosh^2 (2aT) \sinh^2 (2ar)
- 3 \right )}{ \cosh(2aT) \exp \left (
c \cosh^2 \left (2aT \right ) \sinh^2 \left (2ar \right ) \right )},
\\
p = \rho + \frac{8a^2}{\cosh(2aT) \exp \left (
c \cosh^2 \left (2aT \right ) \sinh^2 \left (2ar \right ) \right )}. 
\hspace{10mm}
\end{eqnarray*}
which are now obviously non-negative (provided $c \geq 3/4$ as stated above).

The kinematical quantities of this solution are
\begin{eqnarray*}
\theta = \frac{a \sinh (2aT) \left ( 2 c \cosh^2 (2aT) \sinh^2 (2ar) 
- 3 \right )}{ \sqrt{\cosh (2aT)} \exp \left (
\frac{c}{2} \cosh^2 \left (2aT \right ) \sinh^2 \left (2ar \right ) \right )},
\\
\mbox{\boldmath$a$}=  a c \cosh^2 (2aT) \cosh(4ar) 
\mbox{\boldmath$dr$}. \hspace{7mm}
\end{eqnarray*}
This solution is Petrov type I everywhere (provided $c\neq 0$)
except on the axis of symmetry $r=0$ where
it degenerates to type D as requires the axial symmetry. 

From the expressions above we see that both the density and pressure are
regular everywhere and also the kinematical quantities are non-singular
in the whole spacetime. In fact, it is easy to see that all the Riemann
invariants of this solution are regular everywhere so that this metric contains
no singularity at all. Thus we have found a cosmological solution which
satisfies the Strong Energy Condition everywhere and which contains
no singularity at all. The first solution satisfying energy conditions
and being complete and regular everywhere was found by Senovilla
{\cite{S1}, this solution was generalizad later to a rather large family
in \cite{RS} where diagonal separable metrics in co-moving coordinates
were studied. Some recent work has tried to isolate that family as being
in some sense unique \cite{DP}. However, very recently the author
has found a solution which is also complete and non-singular (obviously
satisfying energy conditions) and such that the line-element is
non-diagonal and with separable metric functions \cite{M1}. The family
that we present in this paper is the first known non-singular model
with non-separable metric coefficients. 

Finally, let us note that
in the particular case
$c=0$ (which does not satisfy the energy conditions) all the kinematical
quantitites of this solution except the expansion vanish identically and the
perfect fluid satisfies a linear equation of state
$p = -5/3 \rho $. Thus, this family
contains a Friedman-Roberson-Walker
model (with a physically unreasonably equation of state) when the constant $c$
vanishes.

\vspace{3mm}

{\bf Solution III}

\vspace{3mm}

The metric functions for this solution read

\vspace{3mm}

\begin{center}

$\displaystyle{ M= 1 + \epsilon e^{-2at} + \beta e^{-6at}, \hspace{1cm}
 N = 1, 
\hspace{1cm} F = \exp \left ( a t +  \epsilon c^2  e^{2 a \left (t+x \right )}
 \right )},$

\vspace{3mm}

$\displaystyle{G= e^{a\left ( t-x \right )}, \hspace{1cm} P=\exp \left (
2 c e^{ax} \right )},$

\vspace{3mm}

\end{center}
with energy density and pressure given by

\begin{center}

\vspace{3mm}

$ \displaystyle{ \rho = \frac{ a^2 \left ( 3 \epsilon e^{4at} + 4 \epsilon
c^2\beta e^{2a \left (
t+x \right )} + 3 \beta \right )}{ 4 \exp \left (
7at + \epsilon c^2 e^{2a \left ( t+ x \right )} \right ) }, \hspace{1cm}
p= \rho + \frac{2 a^2 \beta}{\exp \left (
7at + \epsilon c^2 e^{2a \left ( t+ x \right )} \right ) } }.$
 
\vspace{3mm}

\end{center}

From the form of the function $M$ it follows that the 
coordinate $t$ can take arbitrarily large values
so that the expression for the density clearly 
imposes
$ \epsilon = 1$ in order to fulfil the energy conditions there.
Similarly, the coordinate
$x$ can take arbitrarily large values and, given that the
dominant term
in the density when $x \rightarrow + \infty$ is $\beta c^2
e^{2a \left (t+x \right )}$,
we have to impose $\beta \geq 0$ in order to satisfy the
non-negativity of the energy everywhere. Summing up, we restrict
the constants of the solution to satisfy
\begin{eqnarray*}
\epsilon= +1 , \hspace{1cm} \beta \geq 0
\end{eqnarray*}
so that the density is positive everywhere. Unfortunately, the pressure is always
bigger than the density except for $\beta = 0$
(this particular case is exactly the
solution $\epsilon= +1$, $\beta=0$ in the subcase $R_0=0$
of the Solution I, so that we will not consider it here
any longer). 

The ranges of variation for the coordinates $t$ and $x$ are
\begin{eqnarray*}
- \infty < t < + \infty,  \hspace{1cm} - \infty < x < \infty.
\end{eqnarray*}
The density and pressure diverge when $t$ tends to
$- \infty$ and it is very easy to show that this singularity is at finite
proper time in the past. On the other hand, the density and pressure
tend to zero when the time coordinate tends to $+ \infty$ and it can
be trivially seen that this happens at an infinite proper
time in the future so that this solution presents a big-bang-like singularity
in the past and it is complete in the future. It
is also easy to prove that this solution contains no other singularities
and that is is complete in the coordinate $x$.

The kinematical quantities of the fluid velocity vector read

\vspace{3mm}

\begin{center}

$ \displaystyle{ \theta = \frac{a}{4}  \sqrt{1 + e^{-2at} +
\beta e^{-6at} } \frac{ 4 c^2 e^{2a \left ( t+x \right )} + 6  }{
\exp \left ( \frac{1}{2} at + \frac{c^2}{4} e^{2a \left ( t+ x \right )}
\right ) }, 
\hspace{1cm} 
\mbox{\boldmath$a$}= a c^2 e^{2a \left ( t+ x \right )} 
\mbox{\boldmath$dx$} }. $

\vspace{3mm}

\end{center}
It follows that the expansion is everywhere positive
and it diverges when we approach the big-bang-like
singularity while it tends to zero when $t$ tends to
$+ \infty$. It is interesting to note that the acceleration (and
therefore also the sheartensor) and the Weyl tensor tend to zero
near the big-bang and the density and pressure satisfy a
linear equation of state in the limit near the big bang. In fact,
the asymptotic behaviour of the density and pressure when 
$t \rightarrow - \infty$ is given by
\begin{eqnarray*}
\rho \longrightarrow \frac{3}{4} a^2 \beta e^{-7at}, \hspace{1cm}
p \longrightarrow \frac{11}{4} a^2 \beta e^{-7at}
\end{eqnarray*}
so that near the big bang they satisfy aproximately $p= \frac{11}{3} \rho$.
Consequently, 
the metric is nearly 
homogeneous and isotropic (a nearly Friedman-Robertson-Walker geometry)
in the vicinities of the big-bang and this geometry inhomogenizes and
anisotropizes as the time goes on. In fact, the curvature singularity at
$t= - \infty$ is Ricci dominated in the sense that
\begin{eqnarray*}
\lim_{t \rightarrow - \infty} \frac{C_{\alpha\beta\gamma\delta}
C^{\alpha\beta\gamma\delta}}{R_{\alpha\beta}R^{\alpha\beta}} = 0,
\end{eqnarray*}
where obviously $C_{\alpha\beta\gamma\delta}$ stands for the Weyl tensor
and $R_{\alpha\beta}$ is the Ricci tensor. Furthemore, in the particular case
$\beta= 0$ (which implies the equation of state $p=\rho$) the singularity
at $t= - \infty $ is an {\it isotropic singularity} in the sense that the metric
is conformally related to a metric which is regular at the $t= -\infty$
hypersurface (see {\cite{GW}} and {\cite{WA}} for a precise definition). The
Weyl tensor $C^{\alpha}_{\beta\gamma\delta}$ in the natural coordinate
cobasis $\{ t,x,y,z \}$ is regular but non-vanishing at $t= -\infty $ so that the
so called {\it Weyl tensor hypothesis} is not satisfied. This hypothesis
(see Penrose {\cite{Pe}) assures
that the appropriate thermodynamic boundary
condition for the Universe is that the Weyl tensor should vanish at any initial
singularity. It has been conjectured and partially proved by Tod
(see {\cite{T1}, {\cite{T2}}) that a perfect fluid model
satisfying a barotropic equation of state with an isotropic singularity
and satisfying the Weyl tensor hypothesis is necessarily an exact 
Friedman-Robertson-Walker cosmology everywhere. Thus, the particular solution
$\beta=0$ here presented supports this conjecture due to Tod.

Regarding the Petrov type of this solution, it can be easily seen that it is
Petrov type I everywhere, except when $c$ vanishes. In
this particular case, the acceleration and the shear tensor also vanish
and the density and pressure depend only on $t$ so that they
obey an equation of state. Thus, when $c=0$, the solution
is an exact Friedman-Roberson-Walker geometry.

\vspace{3mm}

{\bf Solution IV}

\vspace{3mm}

The metric coefficients of this solution are given by
\begin{center}

\vspace{3mm}

$\displaystyle{ M= 1 + \epsilon e^{-2at}, \hspace{1cm} N = 1- e^{-6ax},
\hspace{1cm} F = \exp \left ( at + \frac{3}{2} a \left (t + x\right ) + 
\epsilon c^2 e^{2 a \left (t+x \right )} \right )},$

\vspace{3mm}

$\displaystyle{G= e^{a\left ( t-x \right )}, \hspace{1cm} P= \exp \left (
\int{\frac{2ac e^{ax} dx}{ \sqrt{1 - e^{-6ax} }}} \right )
},$

\vspace{3mm}

\end{center}
where $\epsilon$ is a sign and the expressions for the energy density and
pressure are

\vspace{3mm}
\begin{center}
$ \displaystyle{\rho =  \frac{a^2 \left ( 2 \epsilon c^2
e^{4at+2ax}
+9 e^{2at} + 3 \epsilon e^{6ax}\right )}{2  
 \exp \left ( \frac{9}{2}
at + \frac{15}{2} ax + \epsilon c^2 e^{2a \left (t+x \right )} \right )},
\hspace{1cm} p= \rho - \frac{4a^2 }{\exp \left ( \frac{5}{2} at +
\frac{15}{2} ax + \epsilon c^2  e^{2a \left ( t+x \right )} \right )
} }.$

\vspace{3mm}

\end{center}

The range of variation of the coordinate $x$ is 
\begin{eqnarray*}
x > 0
\end{eqnarray*}
so that the positivity condition for the density
when $x \rightarrow + \infty$ implies 
$\epsilon = +1$.
With this value for $\epsilon$ it follows that the
$\rho$ is positive everywhere and
the pressure is also positive
everywhere and satisfies
\begin{eqnarray*}
p \leq \rho. 
\end{eqnarray*}
Thus, this solution satisfies both the Strong and Dominant energy conditions
everywhere.

Given that $\epsilon=+1$ the range of variation for the coordinate $t$
is the whole real line. When this coordinate tends to $- \infty$, both
the density and pressure diverge as
\begin{eqnarray*}
p, \rho \longrightarrow
\frac{3}{2} a^2 e^{-\frac{3}{2} a\left ( 3t + x\right )}.
\end{eqnarray*}
This singularity at $t \rightarrow - \infty$ is at 
a finite proper time in the past and therefore this solution presents
a big-bang-like singularity. On the other hand, when $t$ aproaches $+ \infty$,
the density and pressure tend to zero and the proper time also diverges so that
this solution contains no singularity in the future. Regarding the coordinate
$x$, the proper length diverges when
this coordinate tends to $+\infty$ so that the metric is complete in that
direction. However, the proper length of the apparent singularity at $x=0$ is
finite so that either the solution is truly singular there or we must extend
it beyond this value by means of an appropriate coordinate change.
In order to perform this extension we define new coordinates $X$ and $T$ as
\begin{eqnarray*}
t = \frac{1}{a} \log \left [ \sinh (aT) \right ], \\
x = \frac{1}{3a} \log \left [ \cosh(3aX) \right ],
\end{eqnarray*}
in which the metric takes the form
\begin{eqnarray*}
ds^2 = \sinh(aT) \left [ \sinh^{\frac{3}{2}} (aT) \cosh^{\frac{1}{2}} (3aX)
e^{c^2 \sinh^2(aT) \cosh^{\frac{2}{3}} (3aX)} \left ( -dT^2 + dX^2 \right ) + 
\right .
\\
\left .  \frac{e^{2ac \int{ \cosh^{\frac{1}{3}}(3aX) dX}}}{\cosh^{\frac{1}{3}}
(3aX)} dy^2 + \frac{e^{-2ac \int{ \cosh^{\frac{1}{3}}(3aX) dX}}}{\cosh^{\frac{1}{3}}
(3aX)} dz^2 \right ].
\end{eqnarray*}
This metric is regular at $X=0$ (which corresponds to $x=0$) and
the range of variation of the new coordinate $X$ can be extended to the
whole real line. An analysis of this metric shows that the only singularity
of this extended manifold is the big-bang singularity
(which is now situated at $T=0$) while the solution is complete in the future
and in the spacelike coordinates. The density and pressure in
the new coordinates read
\begin{eqnarray*}
\rho = \frac{ a^2 \left ( 2 c^2 \sinh^4 \left (aT\right ) \cosh^{\frac{2}{3}}
\left (3aX \right ) +9 \sinh^2\left (aT\right ) + 3 \cosh^2\left ( 3aX \right )
\right )}{2 \sinh^{\frac{9}{2}}\left (aT\right ) \cosh^{\frac{5}{2}}\left (3aX
\right )e^{c^2\sinh^2 \left (aT\right ) \cosh^{\frac{2}{3}} \left ( 3aX\right )
}},\\
p = \rho - \frac{4 a^2}{\sinh^{\frac{5}{2}}\left (aT\right ) \cosh^{\frac{5}{2}}\left (3aX
\right )e^{c^2\sinh^2 \left (aT\right ) \cosh^{\frac{2}{3}} \left ( 3aX\right )
}}, \hspace{15mm}
\end{eqnarray*}
which are obviously positive everywhere and satisfying all the energy conditions
at any point of the extended spacetime. Finally, the kinematical quantities 
of the fluid velocity vector are given by

\begin{center}

\vspace{3mm}

$\displaystyle{
\theta = \frac{a \cosh(aT) \left ( 9 + 4c^2 \sinh^2(aT) 
\cosh^{\frac{2}{3}}(3aX) \right )}{4 \sinh^{\frac{9}{4}} (aT) \cosh^{
\frac{1}{4}}
(3aX) e^{c^2 \sinh^2(aT) \cosh^{\frac{2}{3}} (3aX)}} },$

\vspace{3mm}

$\displaystyle{
\mbox{\boldmath$a$}= \frac{a}{4} \frac{\sinh(3aX)}{\cosh(3aX)}
\left (3 + 4c^2 \sinh^2(aT) 
\cosh^{\frac{2}{3}}(3aX) \right )
\mbox{\boldmath$dX$}.
}$

\vspace{3mm}

\end{center}

This solution can be seen of Petrov type I independently of the values of the
arbitrary parameters of the solution and, consequently, this family does
not contain a Friedman-Robertson-Walker model as a particular case.

\vspace{3mm}

\section*{Acknowledgements}

I am very thankful to J.M.M. Senovilla for a careful reading of this typescript
and for his valuable comments and criticisms. I also wish to
thank the {\it Direcci\'o General d'Universitats, Generalitat
de Catalunya}, for financial support.

All the tensors in this paper have been computed with the
algebraic computer programs CLASSI and REDUCE.

\end{document}